\documentclass[aps,prl,twocolumn,showpacs,superscriptaddress]{revtex4}  % for review and submission
\usepackage{graphicx}  % needed for figures
\usepackage{dcolumn}   % needed for some tables
\usepackage{bm}        % for math
\usepackage{braket}
\usepackage{exscale}                  % correct scaling of math-symbols
\usepackage[intlimits]{amsmath}       % improved mathematical typesetting
\usepackage{amsfonts}
\usepackage{amssymb,amscd}
\usepackage{color}
\usepackage{hyperref}
\usepackage{subfigure}
\usepackage[normalem]{ulem}

\newcommand{\mbf}{\mathbf}

\DeclareMathOperator{\sech}{sech}

\begin{document}

\title{Breaking the Fluctuation-Dissipation Relation by Universal Transport Processes} 

\author{Asier Pi\~neiro Orioli}
\affiliation{JILA, NIST, Department of Physics, University of Colorado, Boulder, Colorado 80309, USA}
\author{J\"urgen Berges}
\affiliation{Institute for Theoretical Physics, Heidelberg University, Philosophenweg 16, 69120 Heidelberg, Germany}

\begin{abstract}
Universal phenomena far from equilibrium exhibit additional independent scaling exponents and functions as compared to thermal universal behavior.
For the example of an ultracold Bose gas we simulate nonequilibrium transport processes in a universal scaling regime and show how they lead to the breaking of the 
fluctuation-dissipation relation. As a consequence, the scaling of spectral functions (commutators) and statistical correlations (anticommutators) between 
different points in time and space become linearly independent with distinct dynamic scaling exponents. As a macroscopic signature of this phenomenon we identify a transport peak in the 
statistical two-point correlator, which is absent in the spectral function showing the quasiparticle peaks of the Bose gas.   
\end{abstract}

   
\maketitle

{\it Introduction.---}Universal scaling properties of systems close to equilibrium, as for critical phenomena near phase transitions, represent a cornerstone in the understanding of complex many-body dynamics~\cite{Hohenberg1977a}. Universality implies that a broad class of different systems can show the same properties, captured by universal scaling exponents and functions. For systems far from equilibrium, aspects of universality have also been investigated. Starting with turbulence~\cite{Kolmogorov1941}, there are important topical developments, such as for defect formation~\cite{del2014universality,Navon2015,clark2016}, coarsening~\cite{Bray1994a.AdvPhys.43.357}, driven dissipative dynamics~\cite{Sieberer2013,Navon2018}, or ageing~\cite{Calabrese2005a.JPA38.05.R133}. However, the underlying mechanisms giving rise to universality far from equilibrium are still to a large extent unexplored. 

An important complication arises from the fact that general far-from-equilibrium systems can break the fluctuation-dissipation relation (FDR)~\cite{henkel2011non,Seifert_2012,Baiesi_2013,Puglisi_2017}. As a consequence, the linear response of a system to an external disturbance is no longer determined by its quantum-statistical fluctuations, such that new universal phenomena and additional scaling exponents can arise. Their determination requires knowledge about both response (or so-called spectral) properties as well as statistical correlations between different points in time and space. 

Estimates of nonequilibrium correlations at different times represent a significant computational effort~\cite{Boguslavski18spec}, and their experimental investigation with a combination of spectroscopic and, e.g., weak projective measurements~\cite{KastnerPRA17} is challenging. The rapid progress in developments of platforms using ultracold quantum gases is especially promising for the study of universal nonequilibrium dynamics. Since these tabletop setups can be well isolated from the environment, they offer particularly clean settings. For isolated quantum systems, nonthermal universality classes have been proposed~\cite{Berges:2008wm,Schole:2012kt,Berges:2014bba,Orioli:2015dxa,Schachner17,Karl2017njp,Walz:2017ffj,Chantesana2018} and experimentally discovered recently~\cite{Pruefer2018,Schmiedmayer2018}, focusing so far on equal-time correlations. 

In this Letter we compute universal far-from-equilibrium properties of a Bose gas in three dimensions. We present results for the complete set of spectral and statistical two-times correlation functions in the universal scaling regime, which allows us to establish the role of the FDR. We find that the gas exhibits two distinctive collective components far from equilibrium, characterized by independent dynamic scaling exponents $z$ and $z_c$. The former is encoded in the dispersion $\omega \sim p^z$ between frequency $\omega$ and momentum $p$ for Bogoliubov-like quasiparticles. Most remarkably, these excitations respect the FDR with a thermal equilibrium distribution at all momentum scales even though the system is still far from equilibrium. In particular, this thermal distribution is observed despite the absence of a zero-mode condensate and the existence of nonequilibrium quasiparticle decay rates scaling with time and momentum.

The other scaling exponent $z_c$ governs the self-similar time evolution of a transport peak in the statistical correlation function. The highly occupied transport modes carry conserved particle number towards low momenta. Specifically, their characteristic momenta scale with central time $\tau$ as $p \sim \tau^{-1/z_\text{c}}$ leading finally to condensation. Strikingly, we find that this transport peak is not visible in the spectral function, thus violating the FDR. It is this breaking that leads to the additional exponent $z \neq z_\text{c}$ in contrast to equilibrium scaling.

{\it Spectral and statistical functions far from equilibrium.--}
We investigate an interacting Bose gas described by a Heisenberg field operator $\hat{\psi}(t,\mbf x)$ with~\footnote{We use natural units in which the reduced Planck constant $\hbar = 1$.}
\begin{equation}
	\left(i\partial_t + \frac{\nabla^2}{2m} - g |\hat{\psi}(t,\mbf x)|^2 \right) \hat{\psi}(t,\mbf x) = - h(t,\mbf x),
\label{eq:gpe_h}
\end{equation}
where the external field $h(t,\mbf x)$ will only be used as a numerical tool to study linear response in the system, but has no influence on the dynamics.
The coupling $g=4\pi a/m$ is related to the mass $m$ and scattering length $a$ in the dilute regime, where $n a^3 \ll 1$ with density $n$. 

We consider a class of far-from-equilibrium situations with no condensate in the initial state, but a high occupancy of noncondensed modes up to the healing momentum scale $Q \sim \sqrt{mgn}$. More precisely, the occupancy is encoded in the statistical function~\cite{Aarts:2001qa}
\begin{equation}
F(t,t',\mbf x,\mbf x') \! \equiv \! \frac{1}{2} \langle \{ \hat\psi(t,\mbf x) , \hat\psi^\dagger\!(t'\!,\mbf x') \} \rangle -  \langle  \hat\psi(t,\mbf x) \rangle \langle \hat\psi^\dagger\!(t'\!,\mbf x') \rangle \! ,
\label{eq:statisticalfunction}
\end{equation}
which is the connected part of the field anticommutator expectation value with density $n \equiv F(t,t,\mbf x,\mbf x)$. The occupation number distribution is given by the spatial Fourier transform $F(t,t,p) = \int d^3\! \Delta x\, e^{-i \mbf p \Delta \mbf x} F(t,t,\Delta \mbf x)$ for a homogeneous system with $\Delta \mbf x \equiv \mbf x - \mbf x'$.

The characteristic initial occupancy is taken to be $F(0,0,Q) \sim 1/\sqrt{n a^3}$, which is large for $n a^3 \ll 1$. This property is often encountered following a nonequilibrium instability or quench at earlier times in a variety of many-body systems~\cite{svistunov1991highly,Berges:2014bba,Pruefer2018}.
Here, it may also be understood as the result of performing a strong cooling quench by abruptly removing high-energy modes~\cite{Chantesana2018,Schmiedmayer2018}.
It turns out that the details of such a far-from-equilibrium state do not matter for the subsequent evolution because an effective loss of memory of the initial state takes place which leads to universality~\cite{Boguslavski2014,Orioli:2015dxa}.

In terms of the field operators, the spectral function is given by the commutator expectation value~\cite{Aarts:2001qa}
\begin{align}
	\rho(t,t',\mbf x,\mbf x') \equiv&\, i \langle [\hat{\psi}(t,\mbf x) , \hat{\psi}^\dagger(t',\mbf x') ] \rangle .
\label{eq:commutatorexp}	
\end{align}
This quantity encodes the equal-time commutation relation $\rho(t,t,\mbf x,\mbf x') = i \delta(\mbf x-\mbf x')$, and the retarded propagator is 
$G_R(t,t',\mbf x,\mbf x') = \rho(t,t',\mbf x,\mbf x')\Theta(t-t')$. 
For its computation we use an ensemble of random external fields with
\begin{equation}
	h(t,\mbf x) = h_w(\mbf x)\, \delta(t-t_w),
\label{eq:hdef_xdep}
\end{equation}
where $t_w$ denotes the ``waiting time'' at which the disturbance is applied,
with $\overline{h_w(\mbf x)} = 0 = \overline{h_w(\mbf x)h_w(\mbf y)}$
and $\overline{h_w(\mbf x)h_w^*(\mbf y)} \,=\,  \sigma_h^2\, \delta(\mbf x-\mbf y) $~\cite{Barrat98pre}.
Here the overbar denotes averaging over different random field configurations with $0 < \sigma_h \ll1$.
The spectral function for any time $t>t_w$ can then be obtained in linear response as~\footnote{Using Eq.~(\ref{eq:hdef_xdep}) and expanding to linear order in $h$ one has $\langle\hat\psi(t,\mbf x)\rangle_h\approx \langle\hat\psi(t,\mbf x)\rangle + i \int d\mbf y\,h_w(\mbf y) \langle [\hat\psi(t,\mbf x),\hat\psi^\dagger(t_w,\mbf y)] \rangle$, where we used $\langle[\hat\psi,\hat\psi]\rangle=0$ and $t\geq t_w$. Multiplying with $h^*_w(\mbf x')$ and averaging over random field configurations yields Eq.~(\ref{eq:rhoh}).}
\begin{equation}
	\rho(t,t_w,\mbf x,\mbf x') \stackrel{t>t_w}{=} \lim_{\sigma_h \rightarrow 0} \frac{1}{\sigma_h^2}\, \overline{h_w^*(\mbf x') \langle \hat{\psi}(t,\mbf x) \rangle}_h \, .
\label{eq:rhoh}	
\end{equation}

We consider the spectral function in momentum space and introduce in addition the central and relative time coordinates $\tau \equiv (t+t')/2$ and $\Delta t\equiv t-t'$. The Fourier transform with respect to $\Delta t$ for fixed $\tau$ reads
\begin{equation}
	\rho(\tau,\omega,p) = -i \!\int\limits_{-2\tau}^{2\tau} \! d\Delta t\, e^{i\omega \Delta t} \rho(\tau+\Delta t/2,\tau-\Delta t/2,p) ,
\label{eq:FT_rho_omega}
\end{equation}
where the factor $-i$ is introduced such that $\rho(\tau,\omega,p)$ is real; there is no such factor in the corresponding definition of the real $F(\tau,\omega,p)$.
Alternatively, one may consider a Fourier transform with respect to $\Delta t$ at fixed $t_w$. This provides a very good approximation of the transform at constant $\tau$ for the times we are interested in, and we will employ it in the following instead of Eq.~(\ref{eq:FT_rho_omega}).

\begin{figure}[t!]
\centering
\includegraphics[width=\columnwidth]{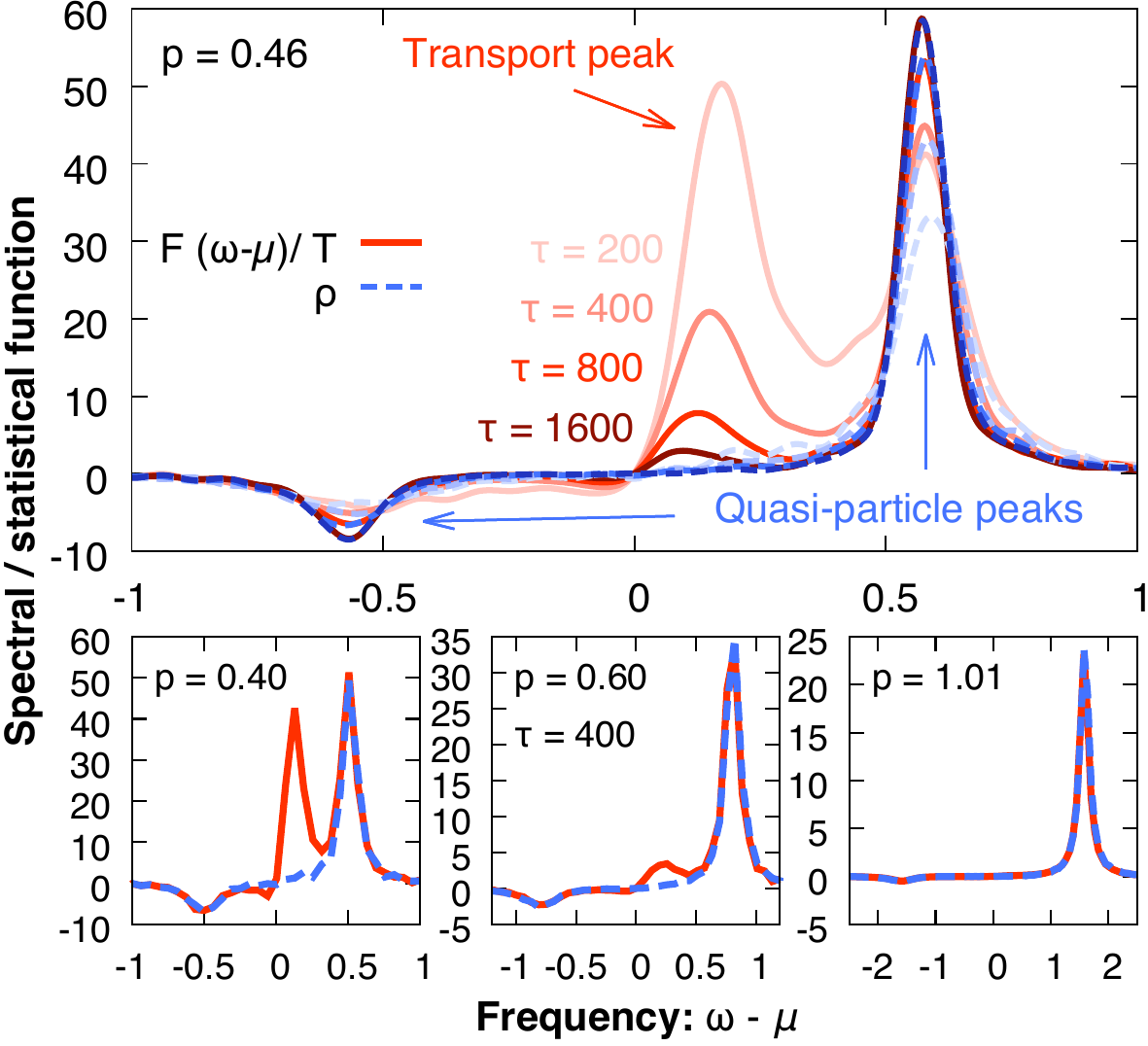}
\caption{Quasiparticle peaks of the spectral function $\rho$ and (rescaled) statistical function $F$ at different times $\tau$ and momenta $p$. The statistical function exhibits an additional transport peak, which is not present in $\rho$, thus breaking the FDR.}
\label{fig:FDR}
\end{figure}

As a consequence of the high occupancies, the system becomes strongly correlated despite being dilute, such that perturbative or mean-field approximations cannot be applied.
In this regime, the quantum dynamics is well approximated in terms of classical-statistical field theory~\cite{TWA}~\footnote{Its range of validity for universal dynamic scaling phenomena is discussed in Ref.~\cite{Boguslavski2014}.}. 
We perform simulations on a 3D lattice with $256^3$ points and lattice spacing $a_s$ such that $Qa_s=1$, where we checked the insensitivity of our infrared results to cutoff changes.
The fields are initialized as $\psi(0,\mbf p)=c_1(\mbf p)+ic_2(\mbf p)$ with Gaussian random numbers $c_i(\mbf p)$, such that $\langle \psi \rangle =0$ and $\langle |\psi(0,\mbf p)|^2 \rangle/V=A/(2mgQ)\Theta(Q-|\mbf p|)$~\footnote{Because $2mgQ\sim\sqrt{na^3}\ll1$ the occupancies are high and the vacuum energy contribution of ``$1/2$'' can be neglected.}, where $A=50$, $V$ is the volume, and $\langle\cdot\rangle$ denotes the average over initial conditions~\cite{Orioli:2015dxa}.
The system is repeatedly evolved using Eq.~(\ref{eq:gpe_h}) as a classical equation with $\hat\psi\rightarrow\psi$ and $h\equiv0$, from which the statistical function is obtained as $F(t,t',p)=\langle\psi(t,\mbf p)\psi^*(t',\mbf p)\rangle/V$.
To get the spectral function we perturb the system with Eq.~(\ref{eq:hdef_xdep}) at the desired time and use Eq.~(\ref{eq:rhoh}).
All results of the simulations are rescaled to represent dimensionless quantities with $F(t,t',p)\rightarrow2mgQ\,F(t,t',p)$ and plotted against rescaled $\tau\rightarrow \tau\,Q^2/(2m)$, $\omega\rightarrow \omega\,(2m)/Q^2$, ${p}\rightarrow {p}/Q$~\footnote{This allows one to scale out $m$ and $g$ of the classical-statistical evolution. The universal character of the physics described further implies insensitivity of our results to variations in $Q$ and $A$.}.

\begin{figure*}[t!]
	\centering
	\includegraphics[width=\textwidth]{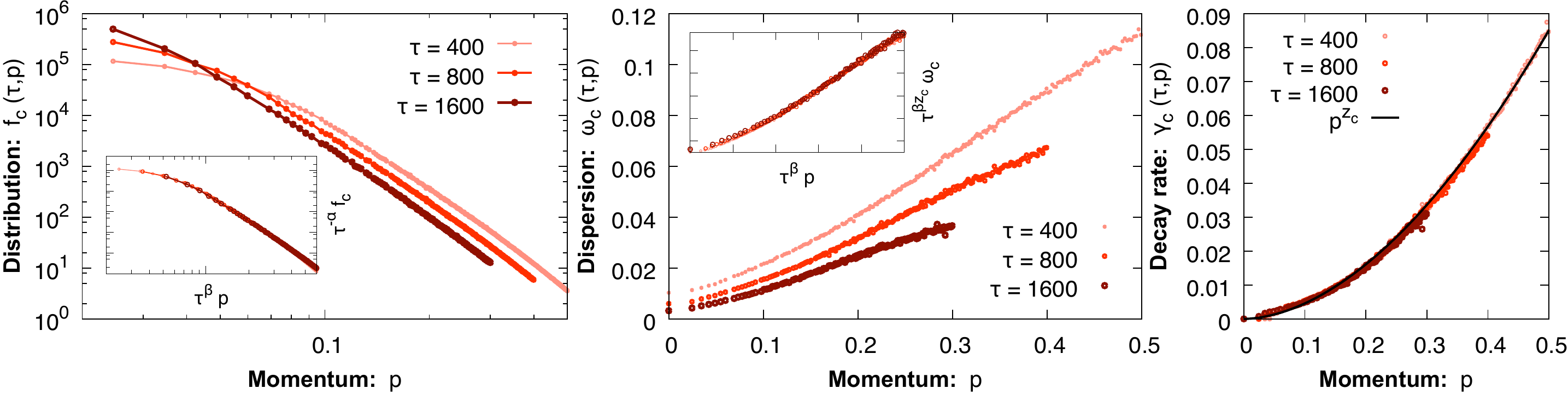}
	\caption{Distribution function $f_\text{c}(\tau,p)$, dispersion $\omega_\text{c}(\tau,p)$ and decay function $\gamma_\text{c}(p)$ for the transport peak. The insets in the left two panels show the rescaled functions to demonstrate their self-similar behavior with universal exponents $\alpha$, $\beta$ and $z_c$.}
	\label{fig:transportpeak}
\end{figure*}

{\it Universal transport peak and breaking of the FDR.---}In thermal equilibrium both $F$ and $\rho$ are related by the FDR, such that their ratio is fixed in terms of the equilibrium temperature $T$ and chemical potential $\mu$ by $F/\rho \stackrel{\rm thermal}{=}1/[\exp\{(\omega-\mu)/T\} -1] \simeq T/(\omega-\mu)$, with the latter equality holding in the classical-statistical regime. Thus we can investigate the role of the FDR by comparing the nonequilibrium time evolution of the rescaled $F(\tau, \omega, p) (\omega-\mu)/T$ and of $\rho(\tau, \omega, p)$. 

The upper plot of Fig.~\ref{fig:FDR} shows the evolution of $F$ (red solid lines) and $\rho$ (blue dashed lines) versus frequency $\omega$ at different times $\tau = 200$ -- $1600$ and fixed momentum $p= 0.46$; the lower plots give the same quantities, however, for fixed time $\tau = 400$ and three different momenta. The spectral function exhibits two characteristic quasiparticle peaks at approximately time-independent frequencies. The statistical function shows two very similar peaks for which the position and amplitude of both $F (\omega-\mu)/T$ and $\rho$ agree remarkably well already at rather early times. However, $F$ exhibits an additional transport peak during the nonequilibrium evolution, which has no counterpart in $\rho$.

More precisely, we find after a short initial period:
\begin{eqnarray}
	\underbrace{F(\tau,\omega,p) \,\simeq\, \frac{T}{\omega-\mu}\, \rho(\tau,\omega,p)} &+& \underbrace{F_\text{c}(\tau,\omega,p) \! { \atop } } . \quad
\label{eq:FDTclassical}
\\
\text{fluctuation-dissipation relation} \! & &  \! \text{transport peak} \nonumber
\end{eqnarray}
The nonequilibrium statistical function approaches rather quickly a form that can be decomposed into a quasiparticle contribution, which fulfills the FDR of thermal equilibrium, and a transport peak contribution $F_\text{c}$ that clearly violates it.
At sufficiently late times the transport peak goes away and the system approaches thermal equilibrium with the FDR fully established. We further emphasize that the transport peak is an emergent phenomenon not present in our Gaussian initial state.

Before the approach to thermal equilibrium, the statistical function is completely dominated by the additional transport peak for low enough momenta $p$. We find the data of $F_\text{c}$ to be well fitted by a hyperbolic secant function of the form
\footnote{We checked that neither a Lorentz nor a Gauss profile can capture the form of the transport peak. The hyperbolic secant in Eq.~(\ref{eq:form_transportpeak}) is an ansatz interpolating between $e^{-[\omega-\omega_c(\tau,p)]^2}$ at small and $e^{-|\omega-\omega_c(\tau,p)|}$ at large $|\omega-\omega_c(\tau,p)|$, as indicated by the numerical data.}
\begin{equation}
	F_\text{c}(\tau,\omega,p) \simeq \frac{2 f_\text{c}(\tau,p)}{\gamma_\text{c}(p)} \sech\!\Big[\{ \omega - \mu - \omega_\text{c}(\tau,p)\}/\gamma_\text{c}(p) \Big]
\label{eq:form_transportpeak}
\end{equation}
with an occupation number distribution $f_\text{c}(\tau,p)$, dispersion $\omega_\text{c}(\tau,p)$, decay function $\gamma_\text{c}(p)$ and the effective
chemical potential $\mu \simeq 1.06$ for the parameters employed. 

The nonequilibrium dynamics of $F_\text{c}$ shows a self-similar scaling behavior associated to the transport of particle number towards lower momentum scales.
The transport phenomenon is encoded in the time evolution of the distribution $f_\text{c}(\tau,p) \equiv \int F_c(\tau,\omega,p) d\omega/(2\pi)$, scaling as~\cite{Orioli:2015dxa}
\begin{equation}
f_\text{c}(\tau,p) = \tau^\alpha f_S(\tau^\beta p) \, 
\label{eq:particlenumber}
\end{equation}
with universal scaling exponents $\alpha$, $\beta$ and scaling function $f_S$ as displayed in the left plot of Fig.~\ref{fig:transportpeak}. The exponent $\beta = 0.55 \pm 0.05$ is to very good accuracy related to $\alpha = 3 \beta$ because of particle number conservation (see also below)~\cite{Orioli:2015dxa}. The scaling function is approximately of the form $f_S(p)\sim 1/[\text{const}+ p^\kappa]$, where $\kappa \simeq 4.5 \pm 0.5$~\cite{Walz:2017ffj,Chantesana2018}. The scaling form [Eq.~(\ref{eq:particlenumber})] entails that a characteristic momentum scales with time as $K(\tau) \sim \tau^{-\beta}$, such that the occupation number of that mode is carried towards low momenta for the positive value of $\beta$.  

\begin{figure}[b!]
	\centering
	\includegraphics[width=\columnwidth]{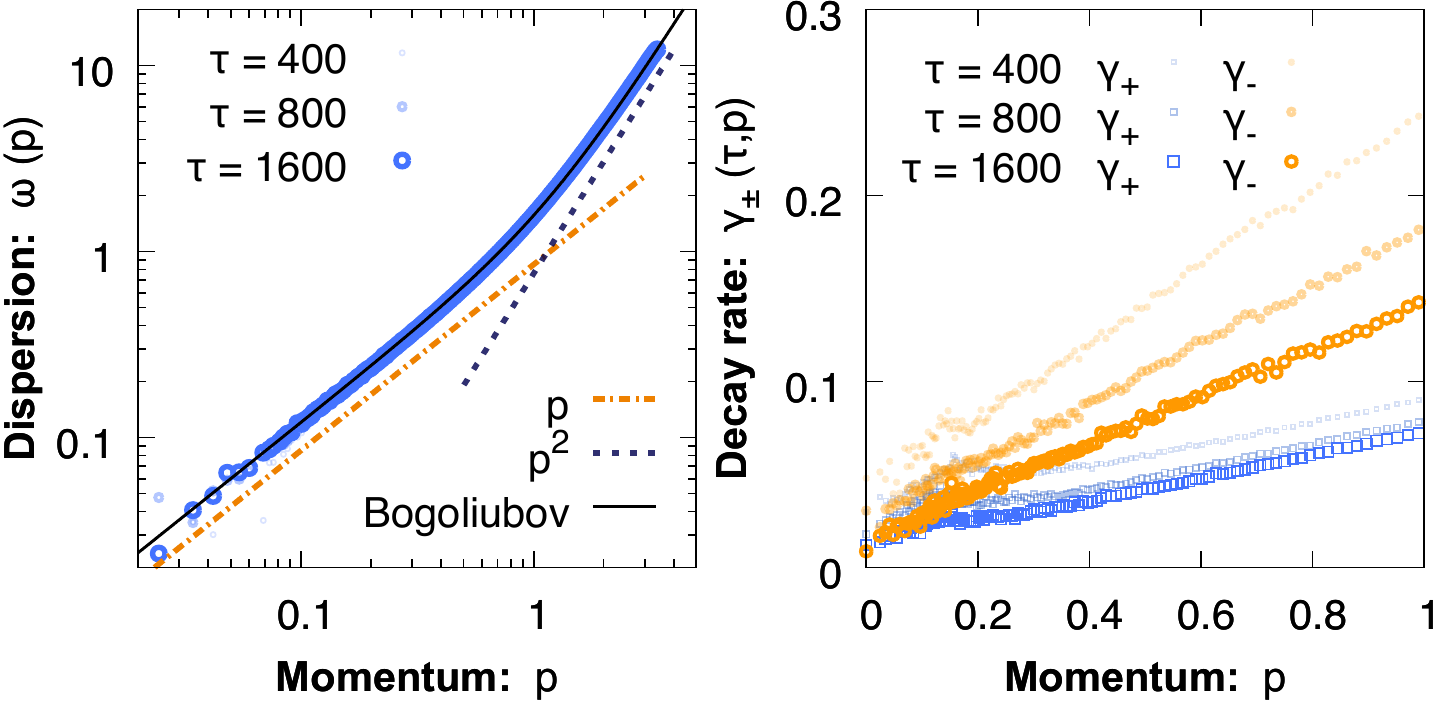}
	\caption{Quasiparticle dispersion relation $\omega(p)$ and time-dependent decay rate $\gamma_\pm(\tau,p)$.}
	\label{fig:quasipart}
\end{figure}

We find that also the dispersion $\omega_\text{c}(\tau,p)$, shown in the middle of Fig.~\ref{fig:transportpeak}, exhibits an emergent scaling behavior:
\begin{equation}
	\omega_\text{c}(\tau,p) = \tau^{-\beta z_c} \omega_S(\tau^\beta p) \, .
\label{eq:dispExtra_approx}
\end{equation}
The scaling exponent $z_c = 1.82 \pm 0.18$ accurately fulfills $z_c = 1/\beta$ (cf.~also~Ref.~\cite{Schachner17}).
Moreover, the right plot of Fig.~\ref{fig:transportpeak} shows the decay function $\gamma_\text{c}(p)$, which follows an approximately time-independent power-law $\gamma_\text{c}(p) \sim p^{z_\text{c}}$. 

{\it Bogoliubov-like quasiparticles out of equilibrium.---}Turning now to the spectral function, we find that rather quickly the data are well fitted by an approximate Lorentzian \footnote{The Lorentzian shape results from an exponentially damped oscillation in the time domain.},
\begin{eqnarray}
	\rho(\tau,\omega,p) &\simeq& \frac{2\, A_+(p)\, \gamma_+(\tau,p)}{\left(\omega-\mu-\omega(p)\right)^2 + \gamma_+(\tau,p)^2} \nonumber\\
&&- \frac{2\, A_-(p)\, \gamma_-(\tau,p)}{\left(\omega-\mu+\omega(p)\right)^2 + \gamma_-(\tau,p)^2}\,.
\label{eq:rho_doublePeak}
\end{eqnarray}
Here, $\mu$ is the same chemical potential as for the statistical function and $\omega(p)$ is the dispersion given in the left plot of Fig.~\ref{fig:quasipart}. 
Even during the far-from-equilibrium stages of the evolution, the dispersion is very well described by the
Bogoliubov form
\begin{equation}
	\omega(p) \simeq \sqrt{\frac{p^2}{2m} \left( \frac{p^2}{2m} + 2gn^T_0 \right)} , 
\label{eq:bog_dispersion}
\end{equation}
assuming at all times a thermal value for the condensate density $n^T_0 \simeq 0.72$ at temperature $T=1.0$ for the parameters employed. 
Accordingly, for small momenta one observes a linear dispersion $\omega(p) \simeq c p$ for sound waves of velocity $c = \sqrt{gn^T_0/m}$. 
The practically time-independent $A_\pm(p)$, with label ``$+$''($-$) for $\omega > \mu$ ($\omega < \mu$), are 
\begin{equation}
	A_\pm(p) \simeq \frac{p^2/2m+gn^T_0 \pm \omega(p)}{2\omega(p)} ,
\label{eq:bog_amplitudes1}
\end{equation}
which at small momenta behave as $A_\pm(p)\sim1/p$.
Therefore, the time dependence of the spectral function is encoded in the decay widths $\gamma_\pm(\tau,p)$ shown in the right plot of Fig.~\ref{fig:quasipart}, which become approximately linear in $p$ at low momenta. This demonstrates the existence of long-lived quasi-particle-like modes also far from equilibrium. 

\begin{figure}[t!]
	\centering
	\includegraphics[width=0.9\columnwidth]{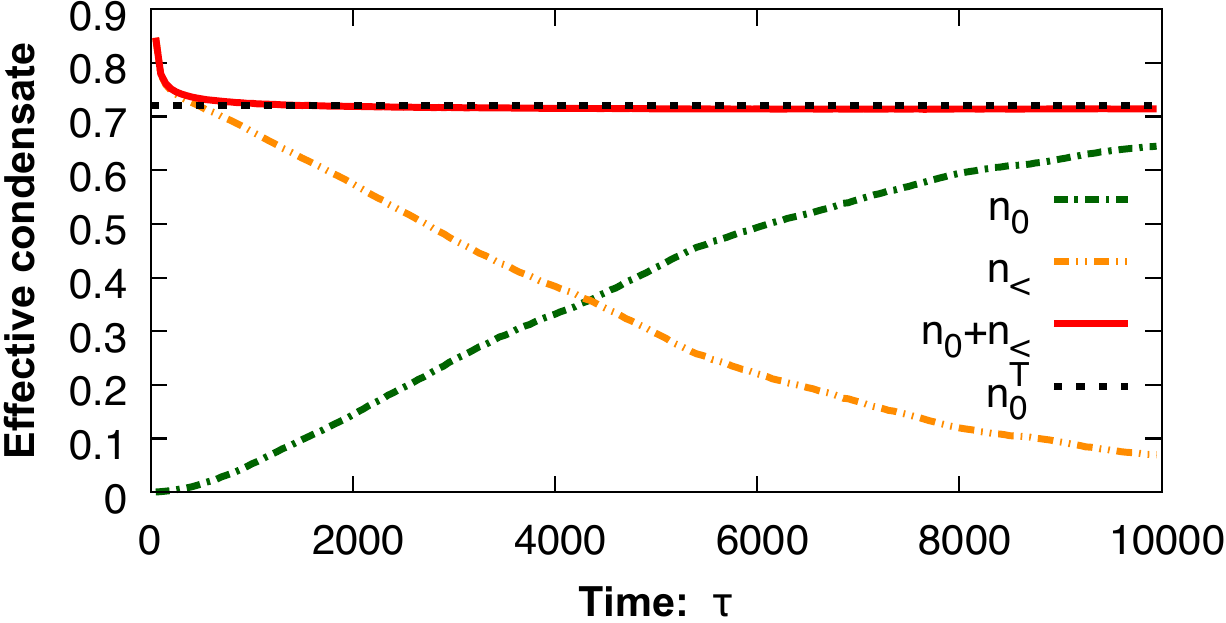}
	\caption{Nonequilibrium zero-mode $n_0(\tau)$ and sum over the dominant transport modes $n_<(\tau)$ defined in Eq.~(\ref{eq:IRsum}), compared to the thermal value  
	of the zero-mode condensate density $n_0^T$.}
	\label{fig:effectivecondensate}
\end{figure}

{\it Effective condensate and unequal-time scaling.} We have established that the Bogoliubov dispersion [Eq.~(\ref{eq:bog_dispersion})], assuming a thermal condensate density $n^T_0$, holds already at early times where the actual condensate density still appears negligibly small. For given volume $V$, the time evolution of the zero-mode $n_0(\tau) = f_\text{c}(\tau,p=0)/V$ is shown in Fig.~\ref{fig:effectivecondensate}. In addition, we display the sum over the dominant transport modes of the distribution function $f_\text{c}(\tau,p)$ up to the time-evolving infrared scale $K(\tau)=K_\text{ref} (\tau/\tau_\text{ref})^{-\beta}$, for $K_\text{ref} = 0.7$, and $\tau_\text{ref} = 400$, as
\begin{equation}
n_<(\tau) = \frac{1}{V} \sum_{0<p\le K(\tau)} f_\text{c}(\tau,p) \, .
\label{eq:IRsum}
\end{equation}
The value of $K_\text{ref}$ is approximately set by the scale below which the transport peak is larger or comparable to the Bogoliubov peaks at time $\tau_\text{ref}$.
Though during the nonequilibrium evolution both $n_0(\tau)$ and $n_<(\tau)$ depend on time, their sum becomes approximately constant. Remarkably, they quickly add up to $n_0(\tau) + n_<(\tau) \simeq n_0^T$, i.e.~the thermal value of the condensate density entering Eq.~(\ref{eq:bog_dispersion}). In this sense, the time-evolving transport peak acts as an ``effective condensate'' for the Bogoliubov modes, allowing the observation of a thermal quasiparticle excitation spectrum in a far-from-equilibrium situation. 

During the build-up of the zero-mode condensate, the statistical and spectral functions exhibit scaling behavior 
\begin{align}
	F(\tau,\omega,p) =&\, \tau^{\alpha + \beta z_c}\, F_S(\tau^{\beta z_c} \omega, \tau^\beta p) \, ,
\label{eq:F_scaling_dyn2}\\
	\rho(\tau,\omega,p) =&\, \tau^{2\beta}\, \rho_S(\tau^{\beta z}\omega,\tau^\beta p) 
\label{eq:rho_scaling_dyn2}
\end{align}
for low enough momenta. In particular, scaling for $\rho$ is only expected in the power-law regime $A_\pm(p)\sim1/p$. 

\begin{figure}[t!]
	\centering
	\includegraphics[width=\columnwidth]{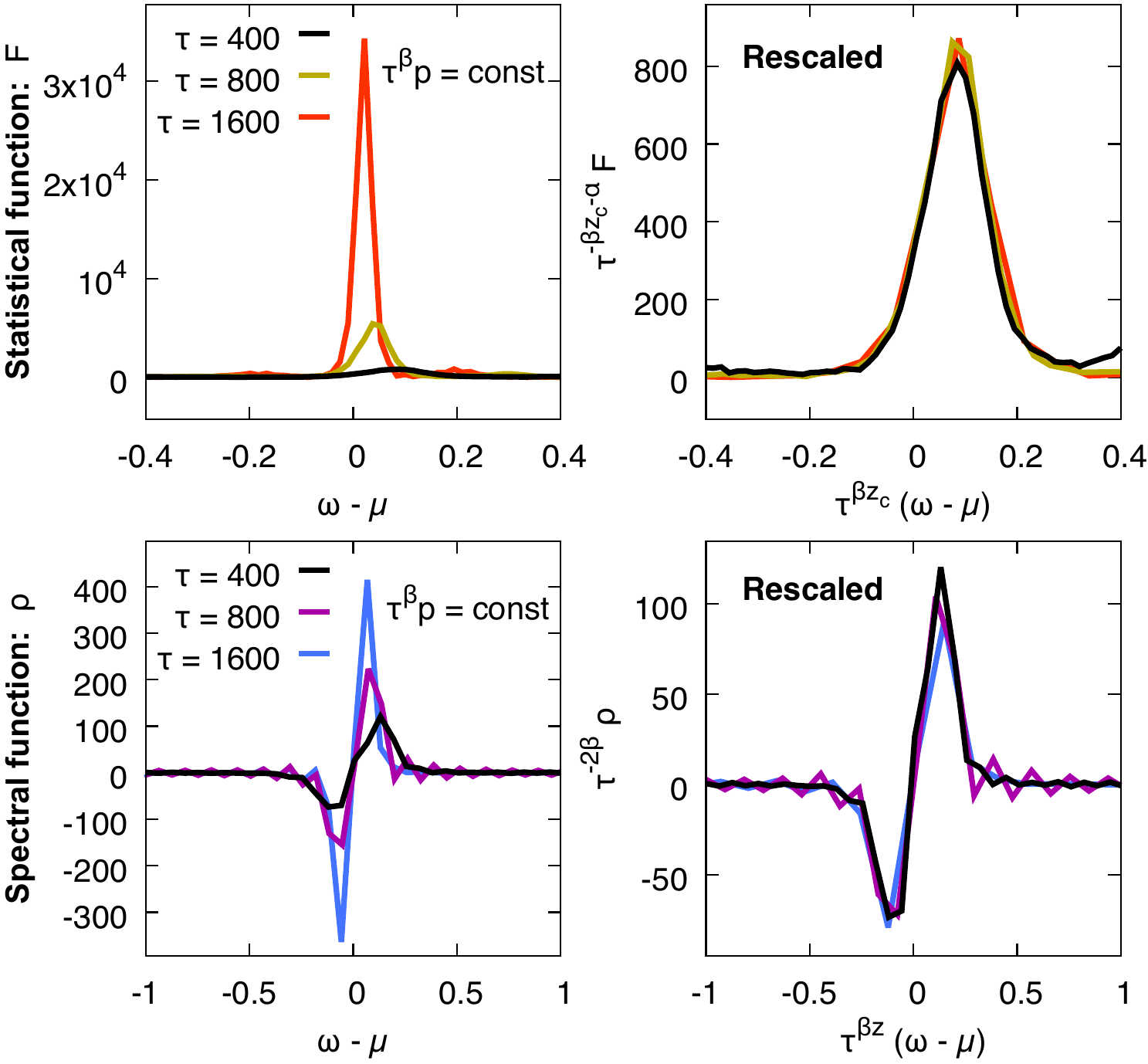}
	\caption{Left panels: $F$ and $\rho$ as a function of $\omega-\mu$ at different $\tau$ keeping $\tau^\beta p$ fixed. Right: Same functions rescaled in time.}
	\label{fig:Frhoscaling}
\end{figure}

The left panels of Fig~\ref{fig:Frhoscaling} show $F$ and $\rho$ as a function of $\omega-\mu$ for three different times and momenta. The right panels display the rescaled functions $\tau^{-\beta z_c-\alpha} F$ versus $\tau^{\beta z_c}(\omega-\mu)$ and $\tau^{-2\beta}\rho$ against $\tau^{\beta z}(\omega-\mu)$. At each time $\tau$, the momentum $p$ is chosen such that the product $\tau^\beta p$ keeps the same value. All curves collapse on top of each other to good accuracy after rescaling, demonstrating the existence of time-independent spectral and statistical nonthermal fixed-point functions  $\rho_S$ and $F_S$.

{\it Conclusions.---}We find two independent dynamic scaling exponents as a consequence of the breaking of the FDR caused by the emergence of a nonequilibrium transport peak. Together with the spectral and statistical fixed-point functions $\rho_S$ and $F_S$, they characterize the far-from-equilibrium universality class of the isolated Bose gas in three spatial dimensions.

The absence of the transport peak in the spectral function preempts effective descriptions such as standard kinetic theory. While this is challenging for theory as well as experimental implementations, our findings provide new possibilities for an efficient characterization of strongly correlated many-body dynamics and the emergence of universality far from equilibrium.
   
We thank K.~Boguslavski, S.~Erne, M.~Oberthaler, M.~Pr\"ufer, T.~Gasenzer, P.~Hauke, A.~M.~Rey, L.~Shen, J.~Schmiedmayer, T.~Zache, V.~Kasper, A.~Schachner, and A.~Schuckert for helpful discussions. This work is part of and supported by the DFG Collaborative Research Centre ``SFB 1225 (ISOQUANT).''


\begin{thebibliography}{99}

	\expandafter\ifx\csname url\endcsname\relax
	  \def\url#1{\texttt{#1}}\fi
	\expandafter\ifx\csname urlprefix\endcsname\relax\def\urlprefix{URL }\fi
	\providecommand{\bibinfo}[2]{#2}
	\providecommand{\eprint}[2][]{\url{#2}}

	\bibitem{Hohenberg1977a}
	\bibinfo{author}{P.~C.~Hohenberg} and \bibinfo{author}{B.~I.~Halperin}.
	\newblock Theory of Dynamic Critical Phenomena.
	\newblock \emph{\bibinfo{journal}{Rev. Mod. Phys.}}
	  \textbf{\bibinfo{volume}{49}}, \bibinfo{pages}{435--479}
	  (\bibinfo{year}{1977}).

	\bibitem{Kolmogorov1941}
	\bibinfo{author}{A.~N.~Kolmogorov}.
	\newblock {The local structure of turbulence in incompressible viscous fluid for very large Reynolds numbers}.
	\newblock \emph{\bibinfo{journal}{Dokl. Akad. Nauk SSSR}} \textbf{\bibinfo{volume}{30}},
	  \bibinfo{pages}{299} (\bibinfo{year}{1941}).
	 
	 \bibitem{del2014universality}
	\bibinfo{author}{A.~Del~Campo} and \bibinfo{author}{W.~H.~Zurek}.
	\newblock Universality of phase transition dynamics: Topological defects from
	  symmetry breaking.
	\newblock \emph{\bibinfo{journal}{Int. J. Mod. Phys. A}}
	  \textbf{\bibinfo{volume}{29}}, \bibinfo{pages}{1430018}
	  (\bibinfo{year}{2014}).

	 \bibitem{Navon2015}
	\bibinfo{author}{N.~Navon},  \bibinfo{author}{A.~L.~Gaunt},  \bibinfo{author}{R.~P.~Smith} and
	  \bibinfo{author}{Z.~Hadzibabic}.
	\newblock Critical dynamics of spontaneous symmetry breaking in a homogeneous Bose gas.
	  \newblock \emph{\bibinfo{journal}{Science}}
	   \textbf{\bibinfo{volume}{347}}, \bibinfo{pages}{167} (\bibinfo{year}{2015}).
	  
	\bibitem{clark2016}
	\bibinfo{author}{L.~W.~Clark}, \bibinfo{author}{L.~Feng} and \bibinfo{author}{C.~Chin}.
	\newblock {Universal space-time scaling symmetry in the dynamics of bosons across a quantum phase transition}.
	\newblock \emph{\bibinfo{journal}{Science}} \textbf{\bibinfo{volume}{354}},
	  \bibinfo{pages}{606} (\bibinfo{year}{2016}).
	  
	  \bibitem{Bray1994a.AdvPhys.43.357}
	\bibinfo{author}{A.~J.~Bray}.
	\newblock Theory of phase-ordering kinetics.
	\newblock \emph{\bibinfo{journal}{Adv. Phys.}} \textbf{\bibinfo{volume}{43}},
	  \bibinfo{pages}{357--459} (\bibinfo{year}{1994}).
	 
	\bibitem{Sieberer2013}
	\bibinfo{author}{L.~M.~Sieberer}, \bibinfo{author}{S.~D.~Huber} , \bibinfo{author}{E.~Altman}  and \bibinfo{author}{S.~Diehl}.
	\newblock Dynamical Critical Phenomena in Driven-Dissipative Systems.
	\newblock \emph{\bibinfo{journal}{Phys. Rev. Lett.}}
	  \textbf{\bibinfo{volume}{110}}, \bibinfo{pages}{195301} (\bibinfo{year}{2013}).

	\bibitem{Navon2018}
	\bibinfo{author}{N.~Navon}, \bibinfo{author}{C.~Eigen},  \bibinfo{author}{J.~Zhang}, \bibinfo{author}{R.~Lopes},
	  \bibinfo{author}{A.~L.~Gaunt}, \bibinfo{author}{K.~Fujimoto},  \bibinfo{author}{M.~Tsubota}, \bibinfo{author}{R.~P.~Smith} and \bibinfo{author}{Z.~Hadzibabic}.
	\newblock {Synthetic dissipation and cascade fluxes in a turbulent quantum gas}.
	  \newblock \emph{\bibinfo{journal}{Preprint at}} \bibinfo{pages}{arXiv preprint arXiv:1807.07564} 	(\bibinfo{year}{2018}).  
	
	\bibitem{Calabrese2005a.JPA38.05.R133}
	\bibinfo{author}{P.~Calabrese} and \bibinfo{author}{A.~Gambassi}.
	\newblock Ageing properties of critical systems.
	\newblock \emph{\bibinfo{journal}{J. Phys. A: Math. Gen.}}
	  \textbf{\bibinfo{volume}{38}}, \bibinfo{pages}{R133} (\bibinfo{year}{2005}).
	
	\bibitem{henkel2011non} 
	\bibinfo{author}{M.~Henkel}, \bibinfo{author}{M.~Pleimling}.
	\newblock Non-Equilibrium Phase Transitions: Volume 2: Ageing and Dynamical Scaling Far from Equilibrium.
	\newblock \bibinfo{publisher}{Springer Netherlands}
	(\bibinfo{year}{2011}).
	
	\bibitem{Seifert_2012}
	\bibinfo{author}{U.~Seifert}.
	\newblock Stochastic thermodynamics, fluctuation theorems and molecular machines.
	\newblock \emph{\bibinfo{journal}{Reports on Progress in Physics}}
	\textbf{\bibinfo{volume}{75}}, \bibinfo{pages}{126001} (\bibinfo{year}{2012}).
	
	\bibitem{Baiesi_2013}
	\bibinfo{author}{M.~Baiesi} and \bibinfo{author}{C.~Maes}.
	\newblock An update on the nonequilibrium linear response.
	\newblock \emph{\bibinfo{journal}{New Journal of Physics}}
	\textbf{\bibinfo{volume}{15}}, \bibinfo{pages}{013004} (\bibinfo{year}{2013}).
	
	\bibitem{Puglisi_2017}
	\bibinfo{author}{A.~Puglisi}, \bibinfo{author}{A.~Sarracino} and \bibinfo{author}{A.~Vulpiani}.
	\newblock Temperature in and out of equilibrium: a review of concepts, tools and attempts.
	\newblock \emph{\bibinfo{journal}{Physics Reports}}
	\textbf{\bibinfo{volume}{709-710}}, \bibinfo{pages}{1 - 60} (\bibinfo{year}{2017}).
	
	\bibitem{Boguslavski18spec} 
	\bibinfo{author}{K.~Boguslavski}, \bibinfo{author}{A.~Kurkela}, \bibinfo{author}{T.~Lappi} and \bibinfo{author}{J.~Peuron}.
	\newblock Spectral function for overoccupied gluodynamics from real-time lattice simulations.
	\newblock \emph{\bibinfo{journal}{Phys.~Rev.~D}} \textbf{\bibinfo{volume}{98}},
	\bibinfo{pages}{014006} (\bibinfo{year}{2018}).

	\bibitem{KastnerPRA17}
	\bibinfo{author}{P.~Uhrich}, \bibinfo{author}{S.~Castrignano},
	\bibinfo{author}{H.~Uys} and \bibinfo{author}{M.~Kastner}.
	\newblock{Noninvasive Measurement of Dynamic Correlation Functions}.
	\newblock \emph{\bibinfo{journal}{Phys.~Rev.~A}} \textbf{\bibinfo{volume}{96}},
	\bibinfo{pages}{022127} (\bibinfo{year}{2017}).

	\bibitem{Berges:2008wm}
	\bibinfo{author}{J.~Berges}, \bibinfo{author}{A.~Rothkopf} and
	  \bibinfo{author}{J.~Schmidt}.
	\newblock {Non-thermal fixed points: Effective weak-coupling for strongly
	  correlated systems far from equilibrium}.
	\newblock \emph{\bibinfo{journal}{Phys. Rev. Lett.}}
	  \textbf{\bibinfo{volume}{101}}, \bibinfo{pages}{041603}
	  (\bibinfo{year}{2008}).

	\bibitem{Schole:2012kt}
	\bibinfo{author}{J.~Schole}, \bibinfo{author}{B.~Nowak} and
	  \bibinfo{author}{T.~Gasenzer}.
	\newblock Critical Dynamics of a Two-dimensional Superfluid near a Non-Thermal
	  Fixed Point.
	\newblock \emph{\bibinfo{journal}{Phys. Rev. A}} \textbf{\bibinfo{volume}{86}},
	  \bibinfo{pages}{013624} (\bibinfo{year}{2012}).

	\bibitem{Berges:2014bba}
	\bibinfo{author}{J.~Berges}, \bibinfo{author}{K.~Boguslavski},
	  \bibinfo{author}{S.~Schlichting} and \bibinfo{author}{R.~Venugopalan}.
	\newblock {Universality far from equilibrium: From superfluid Bose gases to
	  heavy-ion collisions}.
	\newblock \emph{\bibinfo{journal}{Phys. Rev. Lett.}}
	  \textbf{\bibinfo{volume}{114}}, \bibinfo{pages}{061601}
	  (\bibinfo{year}{2015}).

	\bibitem{Orioli:2015dxa}
	\bibinfo{author}{A.~Pi\~{n}eiro Orioli}, \bibinfo{author}{K.~Boguslavski} and
	  \bibinfo{author}{J.~Berges}.
	\newblock {Universal self-similar dynamics of relativistic and nonrelativistic
	  field theories near nonthermal fixed points}.
	\newblock \emph{\bibinfo{journal}{Phys. Rev. D}} \textbf{\bibinfo{volume}{92}},
	  \bibinfo{pages}{025041} (\bibinfo{year}{2015}).
	  
	  \bibitem{Schachner17} 
	\bibinfo{author}{A.~Schachner}, \bibinfo{author}{A.~Pi{\~n}eiro Orioli} and \bibinfo{author}{J.~Berges}.
	\newblock Universal scaling of unequal-time correlation functions in ultracold Bose gases far from equilibrium.
	\newblock \emph{\bibinfo{journal}{Phys.~Rev.~A}} \textbf{\bibinfo{volume}{95}},
	\bibinfo{pages}{053605} (\bibinfo{year}{2017}).
	  
	  \bibitem{Karl2017njp}
	\bibinfo{author}{M.~Karl} and \bibinfo{author}{T.~Gasenzer}.
	\newblock {Strongly anomalous non-thermal fixed point in a quenched two-dimensional Bose gas}.
	\newblock \emph{\bibinfo{journal}{New Journal of Physics}} \textbf{\bibinfo{volume}{19}},
	  \bibinfo{pages}{093014} (\bibinfo{year}{2017}).

	\bibitem{Walz:2017ffj}
	\bibinfo{author}{R.~Walz}, \bibinfo{author}{K.~Boguslavski} and \bibinfo{author}{J.~Berges}.
	\newblock Large-N kinetic theory for highly occupied systems.
	\newblock \emph{\bibinfo{journal}{Phys.~Rev.~D}} \textbf{\bibinfo{volume}{97}},
	\bibinfo{pages}{116011} (\bibinfo{year}{2018}).

	\bibitem{Chantesana2018}
	\bibinfo{author}{I.~{Chantesana}}, \bibinfo{author}{A.~Pi\~{n}eiro Orioli}
	  and \bibinfo{author}{T.~{Gasenzer}}.
	\newblock {Kinetic theory of non-thermal fixed points in a Bose gas}.
	\newblock \emph{\bibinfo{journal}{Preprint at}} \bibinfo{pages}{arXiv preprint arXiv:1801.09490} 	(\bibinfo{year}{2018}).
	
	\bibitem{Pruefer2018}
	\bibinfo{author}{M.~Pr{\"u}fer}, \bibinfo{author}{P.~Kunkel},
	  \bibinfo{author}{H.~Strobel}, \bibinfo{author}{S.~Lannig}, \bibinfo{author}{D.~Linnemann},
	  \bibinfo{author}{C.-M.~Schmied}, \bibinfo{author}{J.~Berges},
	  \bibinfo{author}{T.~Gasenzer} and \bibinfo{author}{M.~K.~Oberthaler}.
	\newblock {Observation of universal dynamics in a spinor Bose gas far from equilibrium}.
	\newblock \emph{\bibinfo{journal}{Nature}} \textbf{\bibinfo{volume}{563}},
	\bibinfo{pages}{217}  (\bibinfo{year}{2018}).

	\bibitem{Schmiedmayer2018}
	\bibinfo{author}{S.~Erne}, \bibinfo{author}{R.~B{\"u}cker}, \bibinfo{author}{T.~Gasenzer}, \bibinfo{author}{J.~Berges} and \bibinfo{author}{J.~Schmiedmayer}.
	\newblock Universal dynamics in an isolated one-dimensional Bose gas far from equilibrium.
	\newblock \emph{\bibinfo{journal}{Nature}} \textbf{\bibinfo{volume}{563}},
	\bibinfo{pages}{225}  (\bibinfo{year}{2018}).

	\bibitem{Aarts:2001qa}
	\bibinfo{author}{G.~Aarts} and \bibinfo{author}{J.~Berges}.
	\newblock Nonequilibrium time evolution of the spectral function in quantum field theory.
	\newblock \emph{\bibinfo{journal}{Phys.~Rev.~D}} \textbf{\bibinfo{volume}{64}},
	\bibinfo{pages}{105010} (\bibinfo{year}{2001}).

	\bibitem{svistunov1991highly}
	\bibinfo{author}{B.~V.~Svistunov}.
	\newblock Highly nonequilibrium Bose condensation in a weakly interacting gas.
	\newblock \emph{\bibinfo{journal}{J. Moscow Phys. Soc}}
	\textbf{\bibinfo{volume}{1}}, \bibinfo{pages}{373} (\bibinfo{year}{1991}).
	  
	\bibitem{Boguslavski2014}
	\bibinfo{author}{J.~Berges}, \bibinfo{author}{K.~Boguslavski},
	\bibinfo{author}{S.~Schlichting}, and \bibinfo{author}{R.~Venugopalan}.
	\newblock Basin of attraction for turbulent thermalization and the range of validity of classical-statistical simulations.
	\newblock \emph{\bibinfo{journal}{Journal of High Energy Physics}}
	\textbf{\bibinfo{volume}{1405}}, \bibinfo{pages}{054} (\bibinfo{year}{2014}).
	  
	\bibitem{Barrat98pre} 
	\bibinfo{author}{A.~Barrat}.
	\newblock Monte Carlo simulations of the violation of the fluctuation-dissipation theorem in domain growth processes.
	\newblock \emph{\bibinfo{journal}{Phys.~Rev.~E}} \textbf{\bibinfo{volume}{57}},
	\bibinfo{pages}{3629} (\bibinfo{year}{1998}).
	
	\bibitem{TWA}
	\bibinfo{author}{A.~Polkovnikov}.
	\newblock Phase space representation of quantum dynamics.
	\newblock \emph{\bibinfo{journal}{Annals Phys.}} \textbf{\bibinfo{volume}{325}},
	\bibinfo{pages}{1790} (\bibinfo{year}{2010}).







	  
	



\end{thebibliography}
\end{document}